\documentclass[12pt,reqno]{amsart}
\usepackage{amsmath,amsfonts,amssymb,amscd,amsthm,amsbsy}
\newtheorem{thm}{Theorem}
\newtheorem{lemma}{Lemma}
\newtheorem{cor}[thm]{Corollary}

\theoremstyle{definition}
\newtheorem*{definition}{Definition}
\newtheorem*{remark}{Remark}
\newtheorem*{remarks}{Remarks}

\makeatletter
\@addtoreset{equation}{section}

\makeatother

\newcommand{\A}{\mathcal{A}}
\newcommand{\G}{\mathcal{G}}
\newcommand{\Q}{\mathcal{Q}}

\def\div{\operatorname{div}}
\def\eps{\varepsilon}

\newcommand{\F}{{\mathcal F}}
\newcommand{\R}{{\mathbb R}}
\newcommand{\Beta}{{B}}

\newcommand{\di}{3}

\newcommand{\ir}{\int_{\R^\di}}
\newcommand{\is}{\int_{S^{2}}}

\newcommand{\ph}{\varphi}

\begin{document}

\author{A. V. Bobylev($^*$), I. M. Gamba($^{**}$) and V. A. 
Panferov($^{\dagger}$)}

\address{\(^*\) Department of Mathematics, Karlstad University, Karlstad, SE-651\,88 Sweden}

\address{\(^{**}\) Department of Mathematics,
The University of Texas at Austin,
Austin, TX 78712-1082 USA}

\address{\(^{\dagger}\) Department of Mathematics,
The University of Texas at Austin, Austin, TX 78712-1082 USA;
current address: Department of Mathematics and Statistics,
University of Victoria, Victoria B.C. V8W 3P4 Canada}

\title{ Moment inequalities and high-energy tails 
for Boltzmann equations with inelastic interactions}

\begin{abstract}
We study the high-energy asymptotics of the steady velocity
distributions for model systems of granular media in various 
regimes. The main results obtained are integral estimates
of solutions of the hard-sphere Boltzmann equations, which 
imply that the velocity distribution functions \(f(v)\) 
behave in a certain sense as \(C\exp(-r|v|^s)\) for \(|v|\) 
large. The values of \(s\), which we call {\em the orders 
of tails}, range from \(s=1\) to \(s=2\), depending on the 
model of external forcing. The method we use is based on
the moment inequalities and careful estimating of constants
in the integral form of the Povzner-type inequalities.  \\ \\
{\sc Keywords:}\!  Boltzmann equation,
inelastic interactions, granular media, 
moments, high-energy tails, Povzner-type inequalities. 
\end{abstract}

\maketitle

\markboth{\sc A.V. Bobylev,  I.M. Gamba and V.
Panferov}
{\sc Estimates of high-energy tails for the
Boltzmann equations}

\vspace{-0.8cm}
\section*{Introduction}

In this paper we address the problem of high-energy 
asymptotics for solutions of kinetic equations 
used for modeling dilute, rapid flows of 
granular media. Granular systems in such regimes 
are interesting from a physical point of view, 
since they show a variety of interesting and 
unexpected properties.
They also appear in a growing number of industrial 
applications. Much of the interest to kinetic 
models in this context comes from the fact that 
such models provide a systematic way of derivation 
of hydrodynamic equations based on the principles 
of particle dynamics. They are also useful for 
numerical modeling of granular flows.
We refer the reader to the 
review papers \cite{Ce,Go1,Go} for a general exposition 
of the subject.

In dilute flows, the binary collisions are often 
assumed to be the main mechanism of particle 
interactions. The effect of such collisions
is modeled by collision terms of the Boltzmann 
or Enskog type. An important feature of 
collisions of granular particles is 
their inelastic character: in each collision 
a certain fraction of the kinetic energy 
is dissipated. This introduces some interesting 
features in the equations: in particular, the 
only functions  on which the collision
operator vanishes are delta functions corresponding to 
all particles at rest (or moving with the same 
velocity).  

To obtain other nontrivial steady states in 
granular systems, as a general rule, a certain 
mechanism of the energy inflow is required. 
Experimentally this can be achieved, for example,
by shaking a 
vessel with granular particles. In terms 
of equations, several simplified models have 
been proposed, in the space-homogeneous case,
which include forcing 
terms of various types \cite{WiMa,VaEr,ErBr2}. 
Examples of such terms are diffusion 
(in the velocity space) and Fokker-Planck 
operators which correspond 
to a physical model of a system of particles 
in a thermal bath. 

Other important types of problems which lead to 
similar equations are related to self-similar 
solutions in the homogeneous cooling problem
and the problem of shear flow 
\cite{EsPo,Ce1,BoGrSp}. In both cases
the equations can be transformed, after an 
appropriate change of variables, to 
space-homogeneous steady problems for the 
Boltzmann-type equations with force terms that 
correspond to the negative or anisotropic 
friction. 

One of the interesting features of granular 
flows, which has been studied actively in the framework 
of kinetic theory, is the the non-Maxwellian 
behavior of the steady velocity distributions. 
In fact, experimental data, theoretical predictions 
and numerical evidence suggest that typical velocity 
distributions in rapid granular flows have high-energy 
asymptotics (or ``tails'') given by the 
``stretched exponentials'' \(\exp(-r|v|^s)\) 
with \(s\) generally not equal to \(2\) (the 
classical, Maxwellian, case), or display 
power-like decay for \(|v|\) large (see 
\cite{GaPaVi,ErBr2} and references therein). 

The precise form of the asymptotics is determined 
by several factors, among which are the details
of the interactions and the forcing models.
In the present paper we study the model 
space-homogeneous system with hard-sphere 
collisions and four types of forcing terms. 
We distinguish between the cases of 
(i) diffusion (Gaussian heat bath), 
(ii) diffusion with friction (Fokker-Planck 
type terms), 
(iii) negative friction (obtained in 
a self-similar transformation in the homogeneous 
cooling problem, and (iv) 
anisotropic friction
which appears in the shear flow 
transformation.

We obtain integral estimates for steady solutions 
using functionals of the form \eqref{expm_rs}, which 
indicate that solutions have high-energy ``tails''
given by the ``stretched exponentials'' \(\exp(-r|v|^s)\), 
with \(s\) depending on the forcing terms. We obtain 
the values \(s=\frac{3}{2}\) for the pure diffusion case, 
\(s=2\) for the diffusion-friction heat bath, 
\(s=1\) for the negative friction case and \(s\ge1\)
in the case of the shear flow. 
Our method is based on representing functionals 
\eqref{expm_rs} in terms of symmetric moments, 
studying the infinite system of inequalities 
satisfied by these moments and using a sharp 
integral form of the so-called Povzner inequalities, 
similar to the one studied by one 
of the authors \cite{Bo} in the case of the 
classical (elastic)
space-homogeneous  Boltzmann equation.
We expect that the estimates obtained 
for the moment inequalities can be used for studying 
the time-dependent moments and the time-evolution of the 
tails, which should be an object of a separate 
study. 

The problem of high-energy tails for solutions 
of the inelastic
hard-sphere Boltzmann with diffusion model has been studied 
previously 
by several authors \cite{EsPo,VaEr,ErBr1,ErBr2} 
by the methods of formal asymptotic 
analysis (a formal argument becomes  
particularly simple if one discards the ``gain'' term in 
the kinetic equations). The general idea that appears 
in those papers is that for functions
of the type \(h(v)=C\exp(-r|v|^s)\), with \(s\le2\), 
the ``gain'' term \(Q^+(h,h)\) in the collision
operator is a small perturbation of the 
loss term for \(|v|\) large. Our approach 
based on studying the moments of the collision 
terms allows us to quantify this idea for the 
solutions of the original problem, without
making apriori assumptions about their 
asymptotic behavior. A rigorous analysis of 
the problem with diffusive forcing has
been performed in \cite{GaPaVi}, where it 
was proved, in particular, that steady solutions are 
infinitely differentiable and decay faster 
than any polynomial for \(|v|\) large. 
A lower bound of the type 
\(C\exp(-r|v|^{3/2})\) was also 
established by using a comparison principle.  
The problem has also been studied numerically 
by a number of authors 
\cite{MoSa,BiShSwSw,MoShSw,GaRjWa}.

Another series of related results was obtained for 
the so-called inelastic Maxwell models 
\cite{BoCaGa,KrBe,KrBe2}, which are approximate 
equations obtained by replacing the collision 
rate in the Boltzmann operator by a relative
velocity independent mean value. 
The equations for the Maxwell models simplify
significantly by using the Fourier transform, 
and the equations for symmetric moments build 
a closed infinite recursive system 
\cite{BoCaGa,KrBe2}. Using the Fourier 
transform methods, Bobylev and Cercignani 
\cite{BoCe1} found solutions to the inelastic 
Maxwell model with a Gaussian heat bath which have
high-energy tails \(\,\exp(-r|v|)\).
In the case of self-similar scaling
for the inelastic Maxwell models, solutions with
power-like tails were found \cite{BaMaPu,ErBr,KrBe1}, 
and it was conjectured by Ernst and 
Brito \cite{ErBr1} that such solutions determine 
the universal long-time asymptotics of the 
time-dependent solutions in the space-homogeneous 
cooling problem. This conjecture has 
recently been proved by Bobylev, Cercignani 
and Toscani \cite{BoCe2,BoCeTo}. 

It is clear, however, that while Maxwell models may give 
reasonable approximations of the macroscopic quantities, 
the details of the velocity distributions can differ 
significantly from the hard-sphere case. In particular, 
this is true with respect of the high-energy asymptotics 
which depends crucially on the behavior of collision rate 
for large relative velocities, as can be easily seen 
from the formal asymptotic arguments of the type 
presented in \cite{ErBr2, EsPo,VaEr}.

On the other hand, 
there is an noticeable gap in the development of rigorous 
mathematical theory, between the Maxwell and hard 
sphere models.  Therefore, the aim of this paper is 
to develop rigorous methods that would allow us to 
study solutions of the hard-sphere Boltzmann equation, 
with a particular attention to the high-energy asymptotics. 

The paper is organized as follows. In Section~\ref{sec:prelim}
we present the problem and formulate the main results. 
One of the the most important technical aspects of our study is 
obtaining a precise integral form of the Povzner-type 
inequalities, which we study in Section~\ref{sec:povz}. 
Section~\ref{sec:mom} is devoted
to the moment inequalities specific to the hard-sphere 
case. 
We formulate the inequalities in terms of the normalized
symmetric moments which appear as the coefficients of
power series expansions of functionals \eqref{expm_rs}. 
We further study the dependence of the inequalities 
on the parameters to find the conditions under which 
the sequences of the normalized moments have geometric growth. 
Finally, Section~\ref{sec:proof} presents the 
proofs of the main theorems. 

Most of our inequalities can be used in the time-dependent 
case, and therefore, we begin the next Section by considering 
the non-stationary Boltzmann equation. 

\section{Preliminaries and main results}
\label{sec:prelim}
We study kinetic models for space-homogeneous granular media, 
in which the one-particle distribution function \(f(v,t)\), 
\(v\in\R^3\), \(t>0\) is assumed to satisfy the following 
equation: 
\begin{equation}
\label{be:force}
\frac{\partial f}{\partial{t}} = \Q(f,f) + \G(f).
\end{equation}
Here \(\Q(f,f)\) is 
the inelastic Boltzmann collision operator, expressing the 
effect of binary collisions of particles, and \(\G(f)\) is 
a forcing term. We will consider three different examples 
of forcing. The first one is the pure diffusion thermal 
bath \cite{WiMa,VaEr,GaPaVi},
in which case
\begin{equation}
\label{force:Gauss}
\G_1(f) = \mu \,\Delta {f},
\end{equation}
where \(\mu>0\) is a constant. The second example is the
thermal bath with linear friction
\begin{equation}
\label{force:FP}
\G_2(f)= \mu \,\Delta {f} + \lambda
\div (v\,f),
\end{equation}
where \(\lambda\) and \(\mu\) are positive constants.

The third example relates to self-similar 
solutions of equation \eqref{be:force}
for $\G(f)=0$ \cite{MoSa,ErBr2}. We denote 
\[
f(v,t) = \frac{1}{v_0^3(t)}\,{\tilde f} \big({\tilde v}(v,t), 
{\tilde t}(t)\big),\quad {\tilde v}=\frac v{v_0(t)},
\]
where
\[
v_0(t) = (a+\kappa t)^{-1}, \quad \ \tilde t(t) =
\frac {1}{\kappa} \ln(1 +\frac\kappa{a} t), \quad a,\,\kappa >0.
\]
Then, the equation for  
$\tilde f (\tilde v, \tilde t)$ 
coincides (after omitting the tildes) 
with equation \eqref{be:force}, where 
\begin{equation}
\label{force:selfsimi}
\G_3(f) = -\kappa \div (v f), \qquad \kappa>0.
\end{equation}
Finally, the last type of forcing is given by the 
term appearing in the shear 
flow transformation (see, for example, \cite{Ce1,BoGrSp})
\begin{equation}
\label{force:Shear}
\G_4(f) = - \kappa v_1
\frac{\partial f}{\partial{v_2}}, 
\end{equation}
where \(\kappa\) is a positive constant.

We assume the granular particles to be perfectly smooth
hard spheres performing inelastic collisions characterized 
by a single parameter: the coefficient of normal 
restitution \(0 < e < 1\). 
To define the collision operator we write
\begin{equation}
\label{co:gain-loss}
\Q(f,f) = \Q^+(f,f) - \Q^-(f,f),
\end{equation}
where the ``loss'' term \(\Q^-(f,f)\) is  
\begin{equation}
\label{co:loss}
\Q^-(f,f) = f(f*|v|),
\end{equation}
and the ``gain'' term \(\Q^+(f,f)\) is most easily defined 
through its weak form: 
\begin{equation}
\label{co:gain}
\ir \Q^+ (f,f)(v)\, \psi(v)\, dv  = 
\frac{1}{4\pi}\ir\ir f(v)\,f(w)\,|u|\, 
\is \psi(v')\,d\sigma\, dw\,dv,
\end{equation}
where \(u=v-w\) is the relative velocity of two particles 
about to collide, and \(v'\) is the velocity after the 
collision. The 
collision transformation that puts \(v\) and \(w\) into 
correspondence with the post-collisional velocities \(v'\) 
and \(w'\) can be expressed as follows:
\begin{equation}
\label{co:transf}
\begin{split}
& v'=v+\frac{\beta}{2}\,(|u|\sigma-u),
\\
& w'=w-\frac{\beta}{2}\,(|u|\sigma-u),
\end{split}
\end{equation}
where we set \(\beta=\frac{1+e}{2}\), and \(0<e<1\) is the 
restitution coefficient. Notice that we always 
have \(\frac{1}{2}<\beta<1\).

Combining \eqref{co:loss} and \eqref{co:gain} and using the 
symmetry that allows us to exchange \(v\) with \(w\) in the 
integrals we obtain the following symmetrized weak form
\begin{equation}\label{co:weak}
\begin{split}
\ir \Q (f,f)(v)\, \psi(v)\, dv  =  \frac{1}{2} 
\ir\ir f(v)\,f(w)\,|u| \A_\beta[\psi](v,w)\, dw\,dv,
\end{split}
\end{equation}
where 
\begin{equation}
\label{coll:psi}
\A_\beta[\psi](v,w) 
= \frac{1}{4\pi}\is(\psi(v')+\psi(w')-\psi(v)-\psi(w))\,d{\sigma}. 
\end{equation}
The weak form \eqref{co:gain} will be sufficient for
the purposes of our study. The usual strong form  
\cite{Ce,Go1,Go} can be obtained from  
\eqref{co:gain} by taking \(\psi(v)=\delta(v-v_0)\) 
(see also \cite{GaPaVi}).  

We will assume that the solutions are
normalized as follows
\begin{equation}
\label{zero_mom}
\ir f(v,t) \, dv = 1, \quad \ir f(v,t)
\,v_i\, dv = 0, \quad i=1,2,3.
\end{equation}

Since the expected behavior of solutions for \(|v|\) large
is \(C\exp(-r|v|^s)\), we  introduce the following functionals:
\begin{equation}
\label{expm_rs}
\F_{r,s}(f) = \ir f(v) \, \exp(r|v|^s)\, dv ,
\end{equation}
and  study the values of \(s\) and \(r\)
for which these functionals are finite. This
motivates the following definition.

\begin{definition}We say that the function \(f\) has an {\em exponential
tail of order \(s>0\)}, if the following supremum
\begin{equation}
\label{tail_temp}
r^*_s = \sup \big\{r>0 \,|\,\F_{r,s}(f) < +\infty \big\}
\end{equation}
is positive and finite.
\end{definition}
In the case \(s=2\) the value of \((r^*_s)^{-1}\)
is known as the {\em tail temperature} of \(f\)
\cite{Bo}.
It is easy to see that the number \(s\) in the above
definition is determined uniquely. Indeed, if for
certain \(s>0\),
\[
0 < r^*_s < + \infty,
\]
then we have
\(
r^*_{s'} = + \infty\),  for every \(s'<s\),
and also
\(
r^*_{s'} = 0\), for every \(s'>s\).

Another useful representation of the functionals
\eqref{expm_rs} is obtained by using the {\em symmetric
 moments} of the distribution function. Setting
\begin{equation}
\label{moments}
m_{p}=\ir f(v) |v|^{2p}\, dv, \quad p\ge 0,
\end{equation}
and expanding the exponential function in
\eqref{expm_rs} into the Taylor series we
obtain (formally)
\begin{equation}
\label{series_rs}
\F_{r,s}(f) = \ir f(v) \,
\Big( \sum\limits_{k=0}^\infty
\,\frac{r^k}{k!} \,|v|^{sk}\Big)\, dv
=  \sum_{k=0}^\infty \,
\frac{m_{\frac{sk}2}}{k!}\, \,r^k.
\end{equation}
Then the value \(r^*_s\) from \eqref{tail_temp}
can be interpreted as the {\em radius of convergence}
of the series \eqref{series_rs}, and the order
of the tail \(s\) is therefore the {\em unique}
value for which the series has a
positive and finite radius of convergence.

We can now formulate the main results of this study.
Our first result concerns the steady states of equation
\eqref{be:force} corresponding to the first three types
of forcing.

\begin{thm}\label{exp_bound:steady}
Let \(f_{i}(v)\), \(i=1,2,3\), be nonnegative steady solutions
of the equations \eqref{be:force}, with the forcing
terms \eqref{force:Gauss}, \eqref{force:FP} and 
\eqref{force:selfsimi},
respectively, and assume that \(f_{i}(v)\)  have
finite moments of all orders. Then \(f_i(v)\)
have exponential tails of orders \(\frac{3}{2}\), \(2\) and \(1\),
respectively.
\end{thm}

For  the shear flow model \eqref{force:Shear},
we obtain the following weaker result.

\begin{thm}\label{exp_bound:steadyshear}
Let \(f_4(v)\) be a nonnegative steady solution
of the shear flow model
\eqref{be:force}, \eqref{force:Shear} that 
has finite moments of all orders. Then the supremum
\(r^*_1\), defined in \eqref{tail_temp}, is finite, and 
therefore, \(s\ge 1\). 
\end{thm}

\begin{remark}
The assumption of finiteness of moments of all orders
is obviously required for the functionals \eqref{expm_rs}
to be finite and for the expansions \eqref{series_rs}
to make sense. However, the moment inequalities we 
establish below also imply the following
{\em apriori estimates} for all cases of the solutions: 
{\em Suppose that a moment \(m_{p_0}\) of any order \(p_0>1\) 
is finite. Then, in fact, all moments are finite and 
the solutions have exponential tails of the corresponding 
order.} This observation is important, since it excludes 
the possibility of power-like decay for solutions of 
the considered equations, as soon as solutions have 
finite mass and finite moment of any order higher 
than kinetic energy.   
\end{remark}

The approach that we take in order to establish the 
above results is based on the moment method, in the 
form developed by one of the authors \cite{Bo}, for 
the classical space-homogeneous Boltzmann equation. 
We study the moment equations obtained by integrating 
\eqref{be:force} against \(|v|^{2p}\):
\begin{equation}
\label{be:mom}
\frac{\partial{m_p}}{\partial t} = Q_p + G_p
\end{equation}
(in the steady case the time-derivative term drops out), where 
\begin{equation}
\label{co-force:mom}
Q_p = \ir \Q(f,f) \,|v|^{2p}\, dv 
\quad\text{and}\quad G_p = \ir \G(f)\, |v|^{2p}\, dv . 
\end{equation}
To investigate the summability of the series 
\eqref{series_rs} we look for estimates 
of the sequence of moments \((m_{p})\), with 
\(p=\frac{sk}{2}\), \(k=0,1,2\dots\), and study the dependence of 
the estimates on \(s\). We will be interested in 
the situation when the series has a finite and 
positive radius of convergence, which means that 
the sequence of the coefficients satisfies
\[
c \,q^k \le \frac{m_{\frac{sk}{2}}}{k!} 
\le C\, Q^k, \quad k=0,1,2...,
\]
for certain constants \(q\) and \(Q>0\).

\section{Povzner-type inequalities for inelastic collisions}
\label{sec:povz}

In this section we establish an important technical result
that will allow us to control the contribution of the 
``gain'' operator in the moment equations \eqref{be:mom}. 
We consider radially symmetric test functions 
\(\psi(v)=\Psi(|v|^2)\).  
The weak form  \eqref{co:gain} of the ``gain'' operator can 
then be written as 
\[
\ir \Q^+ (f,f)(v)\, \Psi(|v|^2)\, dv  =  \frac{1}{2} 
\ir\ir f(v)\,f(w)\,|u|\, \A^+_\beta[\Psi](v,w)\, dw\,dv,
\]
where 
\begin{equation}
\label{Aplus}
\A^+_\beta[\Psi](v,w) = \frac{1}{4\pi}
\is \big( \Psi(|v'|^2) + \Psi(|w'|^2) \big)\, d\sigma. 
\end{equation}

A series of results \cite{Po,El,Gu,De,We,Bo,Lu} obtained in
the case of the classical Boltzmann equation develops
the general idea that for convex functions \(\Psi\) the
expression \eqref{Aplus} is in a certain sense ``smaller''
than the corresponding contribution of the ``loss'' term,
which is given by
\[\A^-[\Psi](v,w) = \Psi(|v|^2)+\Psi(|w|^2).
\]
This type of results is generally known as Povzner-type
inequalities. An approach for obtaining such inequalities
in the case of inelastic collisions has recently been
developed in \cite{GaPaVi}. However, for the purposes
of the present study we need a better control of the
constants in the inequalities than those provided by
the results of \cite{GaPaVi}. We will therefore establish
a sharper version of the Povzner-type inequality for inelastic
collisions, using the ideas of \cite{Bo}. The key point, as in
\cite{Bo}, is to look for estimates of the integral quantity
\eqref{Aplus}, rather than for pointwise estimates of the 
integrand. 

\begin{lemma}
\label{lem:povz}
For every \(\beta\in[\frac{1}{2},1]\) there 
exists a function \(\bar g_\beta(\mu)\), on 
\(\mu\in(-1,1)\), which we define explicitly 
in the course of the proof, such that 
\(\bar g_\beta(\mu)\) is nonnegative, continuous, 
even, nondecreasing for \(\mu\in[0,1]\), 
satisfies
\[
2 \int_0^1 \bar g_\beta(2z-1) \,z \, dz = 1,
\]
and    
\[
\bar g_\beta(\mu) \le 1 + \Big(\frac{1}{\beta}-1\Big)^2,
\]   
and for every smooth function \(\Psi(x)\), defined 
for \(x>0\), nondecreasing and convex, 
\begin{equation*}
\A^+_\beta[\Psi] \le 2
\int_0^{1}  \bar g_\beta(2z-1)\,
\Psi\big(\,z\,(|v|^2+|v_*|^2)\,\big)\, dz\,.
\end{equation*}
\end{lemma}

\begin{remarks}
1) The above inequality is a generalization of 
inequalities (12), (16) from \cite{Bo} to the 
inelastic case, under the extra assumption of 
\(\Psi\) being nondecreasing. Indeed, from the 
conditions on \(\bar g_\beta(\mu)\) it is easy 
to see that in the elastic case \(\beta=1\) we 
must have \(\bar g_1(\mu)=1\). Then the inequality of 
the lemma reduces to 
\[
\A^+_\beta[\Psi] \le 2
\int_0^{1} \Psi\big(z\,(|v|^2+|v_*|^2)\big)\, dz\,,
\]
which is equivalent to the form used in \cite{Bo}. 
2) The smoothness assumption on \(\Psi\) can be 
dispensed with relatively easily, by elaborating 
some of the arguments we use in the proof. On the other 
hand, in the most important examples \(\Psi(x)=x^p\) 
with \(p>1\), which will be used in the moment 
estimates, the needed smoothness is readily available.  
\end{remarks}

\begin{proof} In the proof we use repeatedly the 
following argument: suppose that \(\psi(v)\), 
\(v\in\R^3\) is a convex function. 
Then, for almost every \(a\in\R^3\), and for every
\(b\in\R^3\),
\begin{equation}
\label{sym}
\psi(a+tb)+\psi(a-tb)
\end{equation}
is a nondecreasing function of \(t>0\). (To see 
this differentiate \eqref{sym} in \(t\) and notice 
that a convex function has almost everywhere a 
nondecreasing directional derivative in every 
direction.) 

To apply this 
argument we notice that since \(\Psi(x)\) is 
convex and nondecreasing, then also 
\(\psi(v)=\Psi(|v|^2)\) is convex as a function 
of \(v\in\R^3\). In order to introduce the symmetric 
structure \eqref{sym} into the integrand of 
\eqref{Aplus} we then rewrite 
 $v'$ and $w'$ as
\begin{equation*}
v' = U + \frac{u'}{2}, \quad
w' = U - \frac{u'}{2},
\end{equation*}
where
\(
U = \frac{u+w}{2}
\)
is the velocity of the center of mass, and $u'$ is
the relative velocity after the collision. We then 
have, according to \eqref{co:transf},
\begin{equation}\label{povz4}
u' = (1-\beta) u + \beta |u| \sigma,\
\end{equation}
where \(u\) is the relative velocity before the 
collision. Further, we  pass to the spherical 
coordinates \((\rho, \omega)\) in \(u'\) by setting
\[
u' = \rho |u| \omega, \quad\rho\in\R, \quad \omega \in S^2.
\]
Denoting by \(\nu\) the unit
vector in the direction of \(u\) and  using \eqref{povz4}, 
we can write
\begin{equation}
\label{ons}
\rho\omega = (1-\beta)\nu + \beta \sigma.
\end{equation}

We then perform the change of variables from \(\sigma\) to 
\(\omega\) in the integral \eqref{Aplus}. To do so, notice 
that for every test function \(\ph(k)\), \(k\in\R^3\),
we can formally extend the integration from \(S^2\) 
to \(\R^3\) by writing
\begin{equation}\label{povz8}
\is\ph(\sigma)\,d\sigma 
= \ir\delta\Big(\frac{|k|^2-1}{2}\Big)\,\ph(k)\,dk\,,
\end{equation}
where \(\delta\) is the one-dimensional 
\(\delta\)-function.

Changing variables from $k$ to
\[
k' = (1-\beta)\nu + \beta k, \qquad k'=\rho\omega
\]
and then passing to the spherical coordinates 
\((\rho,\omega)\) we can rewrite the integral 
\eqref{povz8} as 
\begin{equation}
\label{povz9}
\begin{split}
\frac{1}{\beta^3}\is \int_0^\infty \rho^2\, 
\delta\Big(\frac{|\rho\omega-(1-\beta)\nu|^2-\beta^2}{2\beta^2}\Big)\,
\ph\Big(\frac{\rho\omega-(1-\beta)\nu}{\beta}\Big)\,d\rho\,d\omega
\\
= \frac{1}{\beta} \is 
\int_0^\infty \rho^2\,
\delta\Big(\frac{(\rho-a)^2-(a^2+b^2)}{2}\Big)\,
\ph\Big(\frac{\rho\omega-(1-\beta)\nu}{\beta}\Big)\, 
d\rho\,d\omega,
\end{split}
\end{equation}
where 
\begin{equation*}
a=(1-\beta)(\nu\cdot\omega)\quad 
\text{and}\quad  b^2=2\beta-1.
\end{equation*}
The radial integration in \eqref{povz9} can be performed 
explicitly, since
\begin{equation*}
\begin{split}
\int_0^\infty & \rho^2 \,\delta\Big(\frac{(\rho-a)^2 - 
(a^2 + b^2)}{2}\Big) \,d\rho 
= \int_{-a}^\infty \frac{(\rho+a)^2}{\rho} 
\,\delta\Big(\frac{\rho^2 -(a^2 + b^2)}{2}\Big)\,\rho\,d\rho
\\ 
& = \frac{\big(a + \sqrt{a^2 + b^2}\big)^2}{\sqrt{a^2 + b^2}}. 
\end{split}
\end{equation*}
After the radial integration, \(\rho\) in 
\eqref{povz9} will be expressed 
through \(\mu=(\nu \cdot \omega)\) 
according to 
\begin{equation}
\label{def_lambda}
\rho=\lambda(\mu)= (1-\beta)\mu + \sqrt{(1-\beta)^2\mu^2 + 2\beta -1}. 
\end{equation}
(The last equation is nothing but the condition \(|k|^2=1\) expressed 
in the new variables.)
Thus, we obtain the formula
\begin{equation}\label{povz11}
\is\ph(\sigma)\,d\sigma 
= \is \ph\Big(\frac{\lambda\omega-(1-\beta)\nu}{\beta}\Big)
\,g_\beta((\nu\cdot\omega))\,d\omega, 
\end{equation}
where
\begin{equation}\label{povz12}
g_\beta(\mu) = \frac{1}{\beta} 
\frac{\big(a + \sqrt{a^2 + b^2}\big)^2}{\sqrt{a^2 + b^2}} 
= \frac{\lambda^2(\mu)}{\beta(\lambda(\mu)-(1-\beta)\mu)}.  
\end{equation}
Applying identity \eqref{povz11} to \eqref{Aplus} we get
\begin{equation}
\label{Aplus_o}
\A^+_\beta[\Psi] = \frac{1}{4\pi}\is g_\beta(\nu\cdot\omega) 
\Big\{
\Psi\Big(\Big|U+\lambda\frac{|u|}{2}\omega\Big|^2\Big)
+
\Psi\Big(\Big|U-\lambda\frac{|u|}{2}\omega\Big|^2\Big)
\Big\} \, d\omega \, .
\end{equation}

Our next goal is to simplify \eqref{Aplus_o} to get 
a convenient upper bound. 
First, due to the convexity of \(\Psi(|\cdot|^2)\), 
the expression in braces, considered as a function of 
\(\lambda>0\), is monotonically increasing. 
Using \eqref{def_lambda}  it is easy to see that
\begin{equation*}
\label{povz7}
0<2\beta -1 \le \lambda(\mu) \le 1 , 
\end{equation*}
for all \(\mu\in [-1,1]\).  Thus, estimating \(\lambda\) 
by one and setting
\begin{equation}\label{povz13}
E=2U^2+\frac{|u|^2}{2} = |v|^2 + |w|^2,
\end{equation}
we find:
\begin{equation}
\label{int:xi}\begin{split}
\A^+_\beta[\Psi] \le & \frac{1}{4\pi}\is g_\beta(\nu\cdot\omega)
\Big\{
\Psi\Big(E\Big(\frac{1+\xi}{2}\Big)\Big)
+
\Psi\Big(E\Big(\frac{1-\xi}{2}\Big)\Big)
\Big\} \, d\omega,
\end{split}
\end{equation}
where
\begin{equation}\label{povz14}
\xi = \frac{2|U||u|}{E}\,(m\cdot\omega),
\end{equation}
and \(m\) is the unit vector in the direction of \(U\). 

We further symmetrize \eqref{int:xi} by using the change 
of variables \(\omega \mapsto -\omega\). Since the expression 
in braces is invariant under this transformation, we can 
replace the function  \(g_\beta(\mu)\) in  \eqref{int:xi} 
by its symmetrized version
\begin{equation}\label{gbar}
\bar{g}_\beta(\mu)=\frac{1}{2}(g_\beta(\mu)+g_\beta(-\mu)).
\end{equation}
It is now a somewhat tedious but straightforward computation 
to check that $\bar g_\beta(\mu)$ has the properties listed 
in the formulation of the lemma.
Noticing that 
\[
\frac{2|U||u|}{E} \le 1
\]
and using the convexity argument again, this time for 
the function \(\Psi(E(\frac{1+\cdot}{2}))\), we can 
replace \(\xi\) in \eqref{int:xi} by \((m\cdot\omega)\).

Next, we see that, for $|U|$ and $|u|$ fixed, the 
integral \eqref{int:xi} has the structure
\[
\is \ph_1(\nu\cdot\omega)\, \ph_2(m\cdot\omega)\, d\omega\, ,
\]
where $\ph_1$ and $\ph_2$ are nonnegative, even and monotonically
increasing on \([0,1]\). It is easy to show that
the maximum value of the integral is attained when
the vectors \(\nu\) and \(m\) are parallel. 
The integral in \eqref{int:xi} is then bounded by
\begin{equation}
\label{int:nu}
\begin{split}
 \frac{1}{4\pi}\is \bar g_\beta(\nu\cdot\omega)
\Big\{\Psi\Big(E\Big(\frac{1+(\nu\cdot\omega)}{2}\Big)\Big)
+
\Psi\Big(E\Big(\frac{1-(\nu\cdot\omega)}{2}\Big)\Big)\Big\} 
\, d\omega
\\
=\frac{1}{2}\int_{-1}^1 \bar g_\beta(\mu)
\Big\{\Psi\Big(E\Big(\frac{1+\mu}{2}\Big)\Big)
+
\Psi\Big(E\Big(\frac{1-\mu}{2}\Big)\Big)\Big\} \, d\mu\ .
\end{split}
\end{equation}
Using that \(\bar g_\beta\) is an even function of \(\mu\) and 
performing the change of variables \(z=(1+\mu)/2\), we 
arrive at the conclusion of the lemma.  
\end{proof}

In the case when the function $\Psi(x)$ is a power 
function of $x$, the bounds of the lemma take a more 
explicit form, and we obtain the following
important corollary.

\begin{cor}
\label{cor:povz}
Let \(\psi(v)=|v|^{2p}\), where \(p>1\). Then
\(\A_\beta[\psi]\), given by
\eqref{coll:psi}, satisfies the inequality
\begin{equation*}
\A_\beta(|v|^{2p}) \leq \gamma_p (|v|^2 + |w|^2)^p - |v|^{2p} - |w|^{2p},
\end{equation*}
where the constant \(\gamma_p\), defined by \eqref{gamma_p}, 
is strictly decreasing for \(p\ge1\) and 
satisfies \(\gamma_p <\min\{ 1, \frac{4}{p+1}\}\).
\end{cor}
\begin{proof}
Taking \(\Psi(x) = x^p\), we can write 
\[
\A_\beta(|v|^{2p}) \le \A^+_\beta[\Psi] - |v|^{2p} - |w|^{2p}. 
\]
Using Lemma \ref{lem:povz} with \(\Psi(x) = x^p\) we get 
the bound
\[
\A^+_\beta[\Psi] \le \gamma_p  (|v|^2 + |w|^2)^p,
\]
where 
\begin{equation}
\label{gamma_p}
\gamma_p = 2 \int_{0}^1 \bar{g}_\beta(2z-1)\, z^p\, dz. 
\end{equation}
By Lemma \ref{lem:povz}, \(\gamma_1=1\), and so, 
since \(z^p < z\) for all \(z\in (0,1)\), we have
\[
\gamma_p < 1,
\]
for all \(p>1\). On the other hand, estimating 
\(\bar g_\beta(\mu)\) by its maximum 
\[
\bar g_\beta(1) = 1 +  \Big(\frac{1}{\beta}-1\Big)^2 \le 2,
\]
we get 
\begin{equation}
\label{p_ge_3}
\gamma_p \le 4 \int_0^1 z^p\,dz = \frac{4}{p+1}. 
\end{equation}
This completes the proof.   
\end{proof}

\begin{remark}
The expression \eqref{gamma_p} for the constant \(\gamma_p\) 
simplifies in the cases \(\beta=1\) (elastic interactions), 
when 
\[
\gamma_p = \frac{2}{p+1}\,,
\]
and in the case \(\beta=1/2\) (``sticky'' particles), when
\[
\gamma_p =\frac{p2^p+1}{2^{p-2}(p+1)(p+2)}\,.
\]
In the general case the integrand of  \eqref{gamma_p} 
is too complicated to yield an answer in closed form. 
However, the bound \eqref{p_ge_3} is quite useful for 
\(p>3\) and shows the correct ``inverse first power'' 
decay for \(p\) large.
\end{remark}

\section{Moment inequalities}
\label{sec:mom}

The estimate of Corollary \ref{cor:povz} is a crucial step 
to obtaining the moment inequalities in the form 
characteristic for the Boltzmann equation with 
``hard interactions'': \cite{De,Bo}. 
The basic estimate takes a particularly simple 
form when \(p\) is 
an integer, since then the binomial formula can be 
used to obtain the inequality
\[
\A_\beta[|v|^{2p}] + (1-\gamma_p)(|v|^{2p}+|w|^{2p})
\le \gamma_p\sum_{k=1}^{p-1}
{{p}\choose{k}}
|v|^{2k}|w|^{2(p-k)}
\]
(cf. \cite[p. 1189]{Bo}). In the case of non-integer 
\(p\) a similar result can be obtained, after we establish
the following estimates of the binomial expansion.  
\begin{lemma}
\label{lem:binom}
Assume that \(p>1\),  
and let $k_p$ denote the integer part of $\frac{p+1}2$. 
Then for all \(x\), \(y>0\) the following inequalities hold
\begin{equation}\label{povz18}
\begin{split}
\sum_{k=1}^{k_p-1} {{p}\choose{k}}\, 
(x^k y^{p-k} + x^{p-k}y^{k}) & 
\le (x+y)^p-x^p-y^p 
\\
& \le  \sum_{k=1}^{k_p} {{p}\choose{k}}\, 
(x^k y^{p-k} + x^{p-k}y^{k}).
\end{split}
\end{equation}
\end{lemma}
\begin{remarks} 
1) The binomial coefficients for
non-integer \(p\) are defined as 
\[
{{p}\choose{k}} = \frac{p(p-1)\dots(p-k+1)}{k!}, 
\quad k\ge 1; \qquad{{p}\choose{0}} =1. 
\]
2) In the case when \(p\) is an odd integer the first of 
the inequalities becomes an equality which then coincides 
with the binomial formula for \((x+y)^p\). 
\end{remarks}
{\parindent=0pt
\begin{proof}
The proof will be achieved by induction on 
\(n=k_p=1,2,3\dots\) In the case \(k_p=0\) the 
following inequality is satisfied for \(-1<p\le1\):
\begin{equation*}
(x+y)^p-x^p-y^p\le0. 
\end{equation*}
Next, for \(n=1\) and $1<p\le 3$, using 
the above inequality and the identity 
\begin{equation*}
0\le (x+y)^p-x^p-y^p 
= \int_0^x \int_0^y p(p-1) (t+\tau)^{p-2} \,d\tau\,dt ,
\end{equation*}
we obtain
\begin{equation*}
(x+y)^p-x^p-y^p \le 
\int_0^x \int_0^y p(p-1) (t^{p-2}+\tau^{p-2}) \,d\tau\,dt =
 p\,(xy^{p-1}+x^{p-1}y),
\end{equation*}
which provides the base for the induction. 

Assuming now that the inequalities \eqref{povz18} are 
true for \(2n-1< p \le 2n+1\), we write
\begin{equation}\label{povz22}
(x+y)^{p+2} - x^{p+2} - y^{p+2} 
= \int_0^x \int_0^y (p+2)(p+1) (t+\tau)^{p} \,d\tau\,dt .
\end{equation}

By induction hypothesis, the right-hand side of 
\eqref{povz22} is bounded from below by 
\begin{equation*}
\begin{split}
\int_0^x\int_0^y (p+ & 2)(p+1)
(t^p + \tau^p)\,d\tau\,dt 
\\
+ & \int_0^x\int_0^y (p+2)(p+1) 
\sum_{k=1}^{k_p-1} {{p}\choose{k}}\, 
(t^k \tau^{p-k} + t^{p-k}\tau^{k})\,d\tau\,dt ,
\end{split}
\end{equation*}
and from above by a similar expression with \(k_p-1\) 
replaced by \(k_p\). 
Performing the integration, using the identity 
\[
\frac{(p+2)(p+1)}{(k+1)(p-k+1)}{{p}\choose{k}} 
= {{p+2}\choose{k+1}},
\]
and noticing that \(k_{p}+1=k_{p+2}\), we obtain 
the lower bound for \eqref{povz22} in the form 
\[ 
\begin{split}
(p+2)(xy^{p+2}+ x^{p+2}y)
+\sum_{k=1}^{k_p-1} {{p+2}\choose{k+1}} 
( x^{k+1} y^{p+1-k} + x^{p+1-k}y^{k+1} )
\\
=\sum_{k=1}^{k_{p+2}-1} {{p+2}\choose{k}} 
( x^{k} y^{p+2-k} + x^{p+2-k}y^{k} ),
\end{split}
\]
and the upper bound with \(k_{p+2}-1\) replaced by 
\(k_{p+2}\). This completes the induction argument. 
\end{proof}
}

We now establish the following bounds for the moments of the 
collision term \(Q_p\) defined in \eqref{co-force:mom} in 
terms of moments \(m_p\) of the distribution function.

\begin{lemma}
\label{moments:est}
For every \(p>1\),
\begin{equation*}
\begin{split}
-m_{p+\frac{1}{2}}  
\le Q_p 
\le & - ( 1 - \gamma_p)\, m_{p + \frac{1}{2}} 
+ \gamma_p\, S_p \,,
\end{split}
\end{equation*}
where 
\begin{equation}
\label{qp:surplus}
S_p = \sum_{k=1}^{k_p} { {p} \choose{k} }
\big(m_{k+\frac{1}{2}}m_{p-k} + m_{k} m_{p-k+\frac{1}{2}}\big)
\end{equation}
and \(\gamma_p\) is the constant from Corollary \ref{cor:povz}.
\end{lemma}
\begin{proof}
Multiplying the inequality of Corollary \ref{cor:povz} by
\(f(v)f(w)\,|v-w|\)  and integrating with respect to
\(v\) and \(w\), we obtain
\begin{equation}
\label{co-mom:ineq}
\begin{split}
Q_p \le  \frac{\gamma_p}{2} \ir\ir f(v)f(w)\,|v-w|\,
\big((|v|^2+|w|^2)^p - |v|^{2p} - |w|^{2p}\big)\,dv\,dw 
\\
- (1 - \gamma_p) \ir f(v)\,|v|^{2p}\ir f(w)\, |v-w|\,dv\,dw.
\end{split}
\end{equation}
The inner integral in the last term can be estimated as 
\[
\ir f(w)\, |v-w|\,dw \ge |v|.
\]
The last inequality follows by Jensen's inequality, 
since \(f\) is normalized to have unit mass and 
zero mean, and the function \(|v-w|\) is convex 
in \(w\) for every \(v\) fixed.  
Thus, the last integral term in  \eqref{co-mom:ineq} 
is estimated below by 
\[
\ir f(v) \,|v|^{2p+1}\,dv = m_{p+1/2}. 
\]
In the first integral term in  \eqref{co-mom:ineq} we use the 
inequality $|v-w|\le|v|+|w|$ and the upper estimate of 
Lemma \ref{lem:binom} to get 
\begin{equation}
\label{prod:bin}
\begin{split}
|v-w|\,& \big((|v|^2+|w|^2)^p - |v|^{2p} - |w|^{2p}\big)
\\
\le & 
\sum\limits_{k=1}^{k_p} {{p}\choose{k}}
\big(|v|^{2(k+1/2)}|w|^{2(p-k)}+|v|^{2(p-k+1/2)}|w|^{2k}\big).
\end{split}
\end{equation}
Substituting the estimate \eqref{prod:bin} into 
\eqref{co-mom:ineq} and performing the integration we 
obtain the upper bound of the Lemma. 

For the lower bound we use the splitting  
\eqref{co:gain-loss}, neglecting the nonnegative 
\(\Q^+\) term and estimating the moments of  
\(\Q^-\) in the same way as we did for the second 
integral term in \eqref{co-mom:ineq}. This completes 
the proof.  
\end{proof}

We next apply the bounds for the moments of the 
collision terms obtained in Lemma~\ref{moments:est}
to the steady moment equations 
\begin{equation}
\label{mom:bal}
Q_p + G_p=0\,,
\end{equation}
obtained from \eqref{be:mom}.
Under suitable conditions on smoothness and decay for 
\(|v|\) large of the solutions \(f(v)\), the moments 
of the forcing terms are calculated as follows. 
In the case of {pure diffusion} \eqref{force:Gauss} we
have
\begin{equation}
\label{mom:Gauss}
G_p
= \ir f(v) \,\mu \,\Delta |v|^{2p}\, dv
= 2\mu \,p\,(2p+1)\, \, m_{p-1}\, .
\end{equation}
In the case of diffusion with friction \eqref{force:FP}, 
we obtain
\begin{equation}
\label{mom:FP}
\begin{split}
 G_p
= \ir f(v) \,(\,\mu \,\Delta |v|^{2p} -
& \lambda \,v \cdot \nabla|v|^{2p})\, dv
\\
= & - 2 \lambda\, p\, m_{p} + 2\mu\, p\,(2p+1) \, \,m_{p-1} \, .
\end{split}
\end{equation}
Setting $\mu=0$ and \(\lambda=-\kappa\) in the above identity, 
we obtain the case of the self-similar solutions:
\eqref{force:selfsimi}
\begin{equation}
\label{mom:selfsimi}
G_p =  2 \kappa\, p\, m_{p} .
\end{equation}
Finally, for the {shear flow} term \eqref{force:Shear},
we obtain the inequality
\begin{equation}
\label{mom:Shear}
G_p
= 2\kappa \,p \ir f(v)\,v_1\,v_2\, |v|^{2p-2}\,dv \le 2\kappa \,p\,m_p\,.
\end{equation}

Hence, combining the bounds of Lemma~\ref{moments:est} with
the above identities, we find for every \(p>1\),
in the first three cases of forcing
the double inequalities
\begin{equation}
\label{mom-ineq:Gauss}
G_p \le m_{p+\frac{1}{2}} 
\le \frac{1}{1-\gamma_p}\big(G_p+\gamma_p\, S_p\big),
\end{equation}
and in the  case of the shear flow the one-sided inequality
\begin{equation}
\label{mom-ineq:Shear}
 m_{p+\frac{1}{2}} 
\le \frac{1}{1-\gamma_p}\big(2\kappa p\,m_p+\gamma_p\, S_p\big),
\end{equation}
where \(G_p\) are given by \eqref{mom:Gauss}, \eqref{mom:FP} 
and \eqref{mom:selfsimi}, and \(S_p\) is given by \eqref{qp:surplus}. 

Notice that since the terms \(G_p\) and \(S_p\) depend on the 
moments \(m_k\) of order at most \(p\) (\(p-\frac{1}{2}\) in 
the case of \(S_p\) ), inequalities  
\eqref{mom-ineq:Gauss} and \eqref{mom-ineq:Shear} can be 
``solved'' recursively. More precisely, assuming some 
properties of the moments of lower order we can use the 
recursive inequalities to obtain information about the 
behavior of the moments \(m_p\), for \(p\) large.  

In order to study the summability of the series 
\eqref{series_rs}, it is convenient to formulate 
the moment inequalities in terms of the normalized 
moments 
\begin{equation}
\label{mom:c}
z_p = \frac{m_p}{\Gamma(ap+b)},\quad p\ge 0,
\end{equation}
where \(a\) and \(b\) are constants to be determined. 
Indeed, the coefficients of the power series 
\eqref{series_rs} represent a particular case of 
\eqref{mom:c}, with \(p=\frac{sk}{2}\), \(a=\frac{2}{s}\) 
and \(b=1\). We will therefore study the conditions 
on \(a\) and \(b\) under which the sequences of 
normalized moments \(z_p=z_{\frac{sk}{2}}\) have 
geometric (exponential) bounds. 

We will first look for estimates of the term \(S_p\)
in the moment inequalities \eqref{mom-ineq:Gauss} and 
\eqref{mom-ineq:Shear}, expressed in terms of the 
normalized moments \(z_p\). 
We recall the definition of the {Beta} function
\begin{equation}
\label{def:beta}
\Beta(x,y) = \int_0^1 s^{x-1}(1-s)^{y-1}\,ds 
= \frac{\Gamma(x)\Gamma(y)}{\Gamma(x+y)}
\end{equation}
which will be used in the proof of next lemma.

\begin{lemma}
\label{mom:qplus}
Let $m_p = z_p \,\Gamma(ap+b)$ with  \(a\ge1\) and  \(b>0\). 
Then for every $p>1$,
\begin{equation*}
S_p \le A \,\Gamma\big(ap+\frac{a}{2}+2b\big) Z_p,
\end{equation*}
where 
\begin{equation}
\label{Z_p}
Z_p = \max\limits_{1\le k \le k_p} 
\big\{ z_{k+1/2} z_{p-k}, z_k z_{p-k+1/2} \big\}
\end{equation}
and \(A=A(a,b)\) is a constant independent of \(p\). 
\end{lemma}
\begin{proof}
Substituting \eqref{mom:c} in the expression \eqref{qp:surplus}
for \(S_p\) we get
\begin{equation}
\label{sum:Gamma}
\begin{split}
S_p = 
\sum\limits_{k=1}^{k_p} {{p}\choose{k}} \,
\Big( \Gamma\big(ak+\frac{a}{2}+b\big)\,
\Gamma(a(p-k)+b)\, z_{k+1/2}\,z_{p-k} 
\\
+ \,\Gamma(ak+b)\,\Gamma\big(a(p-k)+\frac{a}{2}+b\big)\, 
z_k \,z_{p-k+1/2}\Big)\,. 
\end{split}
\end{equation}
Using \eqref{def:beta}, we can rewrite \eqref{sum:Gamma} as 
\begin{equation}
\label{sum:Gamma2}
\begin{split}
\Gamma(ap+\frac{a}{2}+2b)
\sum\limits_{k=1}^{k_p} {{p}\choose{k}} 
\Big( \Beta\big(ak+\frac{a}{2}+b,a(p-k)+b\big)\, z_{k+1/2}\,z_{p-k} 
\\
+ \,\Beta\big(ak+b,a(p-k)+\frac{a}{2}+b\big)\, z_k \,z_{p-k+1/2}\Big)\,. 
\end{split}
\end{equation}
Next, we estimate the products \(z_{k+1/2}\,z_{p-k}\) 
and \(z_k \,z_{p-k+1/2}\) by their maximum 
\(Z_p\), obtaining the following bound for the sum 
in \eqref{sum:Gamma2}
\begin{equation}
\label{sum:beta}
\begin{split}
Z_p\,\sum\limits_{k=1}^{k_p} {{p}\choose{k}} \,
\Big( \Beta\big(ak+\frac{a}{2}+b,a(p-k)+b\big)
+ \Beta\big(ak+b,a(p-k)+\frac{a}{2}+b\big)\Big)\,
\\
=Z_p\int_0^1 s^{\frac{a}{2}+b-1}(1-s)^{b-1}
\sum\limits_{k=1}^{k_p} {{p}\choose{k}} \,
\big\{ s^{ak}(1-s)^{a(p-k)}
+ s^{a(p-k)}(1-s)^{ak}\big\}\,ds.
\end{split}
\end{equation}
Since the expression in braces depends monotonically on 
\(a\), we estimate it from above
by setting \(a=1\). 
Further, using the lower bound of Lemma 
\ref{lem:binom}, the right-hand side 
of \eqref{sum:beta} is bounded above by
\begin{equation}
\label{bound:beta}
\begin{split}
Z_p \int_0^1 & \Big\{ s^{\frac{a}{2}+ b-1} (1-s)^{b-1}\,
(
1-s^p-(1-s)^p 
)
\\
& + \chi_p\,{{p}\choose{k_p}} s^{\frac{a}{2}+ b-1} (1-s)^{b-1}
\big( s^{k_p} (1-s)^{p-k_p} +s^{p-k_p} (1-s)^{k_p}
\big) \Big\}\,ds \,,
\end{split}
\end{equation}
where \(\chi_p=0\) if \(p\) is an odd integer, 
and \(1\), otherwise. Neglecting the negative terms in 
\(1-s^p-(1-s)^p\) and using the definition of the {Beta} 
function again, we obtain the bound
\begin{equation}
\label{bound:beta1}
\begin{split}
 \Beta & \big(\frac{a}{2}+b,b\big) \\
& + \chi_p\,{{p}\choose{k_p}} 
\Big(\Beta\big(k_p+\frac{a}{2}+b,p-k_p+b\big)+
\Beta\big(k_p+b,p-k_p+\frac{a}{2}+b\big)\Big).
\end{split}
\end{equation}
The first term of \eqref{bound:beta1} is a constant 
independent on \(p\); to estimate the second term 
we recall the following asymptotic formula for the
Gamma functions \cite{AbSt}\,:
\begin{equation}
\label{asympt:Gamma}
\lim_{p\to\infty} \frac{\Gamma(p+r)}{\Gamma(p+s)}\, p^{s-r} = 1 \, ,
\end{equation}
for all \(r\), \(s>0\).
Therefore, taking the first Beta function in the 
second term of \eqref{bound:beta1} for definiteness, 
we obtain 
\[
\begin{split}
{{p}\choose{k_p}} \Beta\big(k_p+\frac{a}{2}+b,p-k_p+b\big)
=  \frac{\Gamma(p+1)\,\Gamma( k_p+\frac{a}{2}+b )\,\Gamma(p-k_p+b)}
{\Gamma( p + \frac{a}{2}+2b )\,\Gamma(k_p+1)\,\Gamma( p-k_p+1)}
\\
\le C\, p^{1-\frac{a}{2}-2b} k_p^{\frac{a}{2}+b-1} (p-k_p)^{b-1}\,.
\end{split}
\]
A similar inequality can be obtained for the other Beta function 
term. 
It is clear now that the second term in \eqref{bound:beta1}
is \(O(p^{-1})\) for \(p\to\infty\), and since it also is locally 
bounded for \(p\ge 0\), it is bounded uniformly in \(p\). 
Denoting now by \(A=A(a,b)\) the uniform bound of 
\eqref{bound:beta1} we obtain the conclusion of the lemma. 
\end{proof}

\begin{remark} A more careful analysis of the expression 
\eqref{sum:beta}  would allow us to obtain 
a sharper upper bound 
\begin{equation}
\label{see_sashas_notes}
C\,p^{-a}Z_p
\end{equation}
for that expression, at least for \(1\le a \le 2\). 
Thus, the factor \(\Gamma\big(ap+\frac{a}{2}+b\big)\) 
in the estimate of the lemma could be improved to 
\(\Gamma\big(ap-\frac{a}{2}+b\big)\). However, 
the result in the present formulation will be sufficient 
to obtain the necessary bounds for the moments, so 
we will not pursue the improved estimates based on
the bound \eqref{see_sashas_notes}.
\end{remark}

We next obtain the simplified inequalities for the 
normalized 
moments \eqref{mom:c}. Substituting \eqref{mom:c} in
the inequalities \eqref{mom-ineq:Gauss} and using
the estimate of Lemma \ref{mom:qplus} we obtain 
in the case of pure diffusion \eqref{mom:Gauss}
\begin{multline}
\label{mineq:z_p1}
{2\mu}\,\frac{\Gamma(ap-a+b)}{\Gamma(ap+\frac{a}{2}+b)}\,
p(2p+ 1)\,z_{p-1} 
\le z_{p+\frac{1}{2}} 
\le  \frac{2\mu}{1-\gamma_p} \, 
\frac{\Gamma(ap -a+b)}{\Gamma(ap+\frac{a}{2}+b)}\,
p(2p+1)\,z_{p-1} 
\\
+ \frac{\gamma_p A}{1-\gamma_p}\, 
\frac{\Gamma(ap+\frac{a}{2}+2b)}{\Gamma(ap+\frac{a}{2}+b)}
\,Z_p\,,\qquad
\end{multline}
for all \(p\ge 1\). In the case of diffusion with 
friction \eqref{mom:FP}, the terms 
\begin{equation}
\label{mineq:xtra1}
-{2\lambda}\,
\frac{\Gamma(ap+b)}{\Gamma(ap+\frac{a}{2}+b)}\,p\,z_{p}
\qquad\text{and}\qquad
-\frac{2\lambda}{1-\gamma_p}\,
\frac{\Gamma(ap+b)}{\Gamma(ap+\frac{a}{2}+b)}\,p\,z_{p}
\end{equation}
will be added to the left and the right-hand sides of 
\eqref{mineq:z_p1}, respectively. For the shear flow 
case \eqref{mom:Shear} we obtain
\begin{equation}
\label{mineq:Shear}
z_{p+\frac{1}{2}} \le \frac{2\kappa}{1-\gamma_p}\,
\frac{\Gamma(ap+b)}{\Gamma(ap+\frac{a}{2}+b)}\,p\,z_{p}
+ \frac{\gamma_p A}{1-\gamma_p}\, 
\frac{\Gamma(ap+\frac{a}{2}+2b)}{\Gamma(ap+\frac{a}{2}+b)}
\,Z_p\,. 
\end{equation}

Using Corollary \ref{cor:povz}, for every \(\eps>0\) and 
for all \(p>1+\eps\), the constants involving \(\gamma_p\) 
can be estimated  as follows:
\begin{equation}
\label{K_eps1}
1\le \frac{1}{1-\gamma_p} \le \frac{1}{1-\gamma_{1+\eps}} = K_{\eps}
\end{equation}
and 
\begin{equation}
\label{K_eps2}
\frac{\gamma_p}{1-\gamma_p} \le \frac{4K_{\eps}}{p+1}.  
\end{equation}
Further, using the identities
\[
z\,\Gamma(z)=\Gamma(z+1)
\quad\text{and}\quad 
z\,(z+1)\,\Gamma(z)=\Gamma(z+2)
\]
and estimating 
\[
0< c_3
\le \frac{2p\,(2p+1)}{(ap-a+b)(ap+1-a+b)} 
\le C_3,
\]
and 
\[
ap+\frac{a}{2}+2b-1 \le C_4\,\frac{p+1}{4},
\]
we can reduce the inequalities \eqref{mineq:z_p1} to 
\begin{equation}
\label{mineq:z_p2}
\begin{split}
{c_3\,\mu}\,\frac{\Gamma(ap-a+b+2)}{\Gamma(ap+\frac{a}{2}+b)}\,z_{p-1} 
\le z_{p+\frac{1}{2}} 
\le  {C_3\, K_\eps\,\mu} \, 
\frac{\Gamma(ap -a+b+2)}{\Gamma(ap+\frac{a}{2}+b)}\,z_{p-1} 
\\
+\, C_4\, K_\eps\, 
\frac{\Gamma(ap+\frac{a}{2}+2b-1)}{\Gamma(ap+\frac{a}{2}+b)}
\,Z_p\,.
\end{split}
\end{equation}
For the additional terms \eqref{mineq:xtra1}
appearing in the equation with friction, we 
use the inequalities
\[
c_5 \le \frac{2p}{ap+b} \le C_5
\]
to estimate them as
\begin{equation}
\label{mineq:xtra2}
-{C_5K_\eps\lambda}\,
\frac{\Gamma(ap+b+1)}{\Gamma(ap+\frac{a}{2}+b)}\,z_{p}
\qquad\text{and}\qquad
-c_5\lambda\,
\frac{\Gamma(ap+b+1)}{\Gamma(ap+\frac{a}{2}+b)}\,z_{p}.
\end{equation}
Finally, for the self-similar solution case we obtain the
inequalities
\begin{equation}
\label{mineq:Shear2}
\begin{split}
c_5\,\kappa\,
\frac{\Gamma(ap+b+1)}{\Gamma(ap+\frac{a}{2}+b)}\,z_{p}
\le z_{p+\frac{1}{2}} \le C_5\,K_\eps\,\kappa\,
\frac{\Gamma(ap+b+1)}{\Gamma(ap+\frac{a}{2}+b)}\,z_{p}
\\
+ C_4\,K_\eps\,
\frac{\Gamma(ap+\frac{a}{2}+2b-1)}{\Gamma(ap+\frac{a}{2}+b)}
\,Z_p\,,
\end{split}
\end{equation}
the last of which is also true in the shear flow case. 

We now study the inequalities 
\eqref{mineq:z_p2}--\eqref{mineq:Shear2} 
for the values of \(a=\frac{2}{s}\) corresponding to 
the proposed orders of tails of the solutions. 
In the case of pure diffusion we take \(a=\frac{4}{3}\) 
and the inequalities \eqref{mineq:z_p2} take the form
\begin{equation}
\label{mineq:z_p3}
c_3\,\mu\,z_{p-1} \le z_{p+\frac{1}{2}} 
\le C_3\, K_{\eps}\, \mu \,z_{p-1} 
+ C_4\,K_{\eps}\,\frac{\Gamma(\frac{4}{3}\,p+\frac{2}{3}+2b-1)}
{\Gamma(\frac{4}{3}\,p+\frac{2}{3}+b)}\,Z_p\,,
\end{equation}
for \(p>1+\eps\). We notice that if \(b<1\), the asymptotic 
formula \eqref{asympt:Gamma} allows us to control the factor 
in front of the \(Z_p\) term in \eqref{mineq:z_p3} in 
the following way: 
\begin{equation}
\label{igamma1}
C_4\,K_{\eps}\,\frac{\Gamma(\frac{4}{3}\,p+\frac{2}{3}+2b-1)}
{\Gamma(\frac{4}{3}\,p+\frac{2}{3}+b)} \le \frac{1}{2}, 
\quad\text{for}\quad p \ge p_1,
\end{equation}
if we take \(p_1\) sufficiently large. 
Inequality \eqref{mineq:z_p3} then becomes
\begin{equation}
\label{mineq:z_p4}
c_3\,\mu\,z_{p-1} \le z_{p+\frac{1}{2}} 
\le C_3\, K_{\eps}\, \mu \,z_{p-1} 
+ \frac{1}{2}\,Z_p,
\quad\text{for}\quad p \ge p_1.
\end{equation}

In the case of diffusion with friction the choice 
\(a=1\) gives us the inequalities
\begin{equation}
\label{mineq:z_p5}
\begin{split}
-C_5\,K_\eps\,\lambda\, z_p + c_3\,\mu\,z_{p-1} 
\le \frac{\Gamma(p+\frac{1}{2}+b)}{\Gamma(p+1+b)}\, 
\, & z_{p+\frac{1}{2}} 
\le -c_5 \lambda \,z_p
\\
+ \, & C_3\,K_\eps\,\mu \,z_{p-1} + C_4\,K_\eps
\frac{\Gamma(p-\frac{1}{2}+2b)}{\Gamma(p+b+1)}\,Z_p\,.
\end{split}
\end{equation}
Taking now \(b<\frac{3}{2}\) and choosing \(p_1\) 
large enough, we obtain using \eqref{asympt:Gamma},
\begin{equation}
\label{igamma2}
C_4\,
\frac{\Gamma(p-\frac{1}{2}+2b)}{\Gamma(p+b+1)} 
\le \frac{c_5 \lambda}{2}
\quad\text{and}\quad
\frac{\Gamma(p+\frac{1}{2}+b)}{\Gamma(p+1+b)} \le 1, 
\quad\text{for}\quad p \ge p_1.
\end{equation}
We can then use \eqref{mineq:z_p3} to obtain 
the following simple inequalities
\begin{equation}
\label{mineq:z_p6}
C_5\,K_\eps\,\lambda\, z_p \ge  c_3\,\mu\,z_{p-1} - \,z_{p+\frac{1}{2}} 
\end{equation}
and 
\begin{equation}
\label{mineq:z_p7}
c_5 \,\lambda \,z_p
\le  C_3\,\mu \,z_{p-1} + \frac{1}{2}\,c_5 \,\lambda\,Z_p\,,
\end{equation}
for all \(p\ge p_1\).  

Finally, in the case of self-similar solutions 
we take \(a=2\), and \eqref{mineq:Shear2} becomes
\begin{equation}
\label{mineq:Shear3}
c_5\,\kappa\,z_{p} \le 
z_{p+\frac{1}{2}} \le C_5\,K_\eps\,\kappa\,z_{p}
+ C_4\,K_\eps\,
\frac{\Gamma(2p+2b)}{\Gamma(2p+b+1)}
\,Z_p\,. 
\end{equation}
We then take \(b<1\) and choose \(p_1\) large enough to 
obtain
\begin{equation}
\label{gamma3}
C_4\,K_\eps\,
\frac{\Gamma(2p+2b)}{\Gamma(2p+b+1)} \le \frac{1}{2},
\quad\text{for}\quad p \ge p_1.
\end{equation}
Inequality \eqref{mineq:Shear3} then simplifies to 
\begin{equation}
\label{mineq:Shear4}
c_5\,\kappa\,z_{p} \le
z_{p+\frac{1}{2}} \le C_5\,K_\eps\,\kappa\,z_{p}
+ \frac{1}{2}
\,Z_p\,, 
\quad\text{for}\quad p \ge p_1.
\end{equation}
The second of these inequalities is also satisfied in 
the shear flow case. 

We have now obtained inequalities for the normalized 
moments \eqref{mom:c} in the form which is simple enough 
to be analyzed and which, as we will see below, will allow
us to prove the results about the tail behavior stated
in Section \ref{sec:prelim}. Abstracting now from the 
precise meaning of the terms in inequalities 
\eqref{mineq:z_p4}, \eqref{mineq:z_p6}, \eqref{mineq:z_p7}
and \eqref{mineq:Shear4} we can say that they express 
the balance between the ``loss terms''
(moments of order \(p+\frac{1}{2}\)), ``gain terms'' 
(terms involving \(Z_p\)), diffusion (moments 
of order \(p-1\)) and  friction or the 
force terms in the shear flow (moments 
of order \(p\)). We notice also that inequalities 
in the form \eqref{mineq:z_p2}, \eqref{mineq:xtra2} 
and \eqref{mineq:Shear2} together with the asymptotic 
formula \eqref{asympt:Gamma} can be used to actually 
{\em derive} the values of \(a\) for which the 
series \eqref{series_rs} has finite and positive 
radius of convergence. For the sake of simplicity, 
since these values are already known from the 
formal arguments, we will not perform these 
computations here. 

\section{ Proofs of Theorems \ref{exp_bound:steady} and 
                             \ref{exp_bound:steadyshear} }
\label{sec:proof}

\begin{proof}[Proof of Theorem~\ref{exp_bound:steady}.]
We will establish  the following statement that will imply 
the conclusion of the Theorem (see also the Remark that 
follows Theorem~\ref{exp_bound:steadyshear}). 
We show that for every
\(p_0>1\) there are positive constants \(c\), \(q\), 
depending on \(m_0\) and \(m_1\) only, and \(C\), \(Q\), 
depending on \(m_0\), \(p_0\) and  \(m_{p_0}\) 
only, such that
\begin{equation}
\label{ineq:m_p}
c \,q^k \le \frac{m_{\frac{sk}{2}}}{k!} \le C\, Q^k, 
\end{equation}
for all \(k \ge \frac{2}{s}\), where \(s=\frac{3}{2}\) in the 
case of the pure diffusion, \(s=2\) in the case of
diffusion with friction, and \(s=1\) for the 
self-similar solutions \eqref{force:selfsimi}. 
Equivalently, we can set
\(a=\frac{4}{3}\), \(a=1\) and \(a=2\) in the respective cases 
and look for the estimates 
\begin{equation}
\label{geom:z_p}
c \,q^p \le z_p \le C\, Q^p,
\end{equation}
for all \(p\ge1\), with \(z_p\) defined as 
in \eqref{mom:c}. (The constants in \eqref{geom:z_p} have 
to be modified to match those in \eqref{ineq:m_p}.) 

Notice that it would be sufficient to prove \eqref{geom:z_p} 
for a {\em certain} value of \(b>0\) in the definition of 
\(z_p\) \eqref{mom:c}.
Indeed, since 
\[
C_1\, p^{b_1-b_2} 
\le \frac{\Gamma(ap+b_1)}{\Gamma(ap+b_2)} 
\le C_2\, p^{b_1-b_2},
\]
changing the value of \(b\) in \eqref{mom:c} essentially 
results in the multiplication of \(z_p\) by the factor 
\(C p^{b_1-b_2}\), which can be compensated for by 
adjusting the constants in \eqref{geom:z_p}. We 
fix the value of \(b<1\) so that inequalities 
\eqref{mineq:z_p3}, \eqref{mineq:z_p4} and 
\eqref{mineq:z_p5} are available for \(p\) 
sufficiently large.

The proof of the inequalities \eqref{geom:z_p} will be 
accomplished in two steps. The first one will be 
to show that \eqref{geom:z_p} holds on the initial interval,  
\(1\le p \le p_1\), where \(p_1\) (dependent on \(p_0\) 
and \(b\)) is chosen so that inequalities \eqref{igamma1} 
and \eqref{igamma2} hold with \(\eps=\frac{1-p_0}{2}\).  

{\em Step 1: Initial interval.} 
We notice that for \(1\le p \le p_1\), the Gamma 
function is bounded both from above and from below:
\begin{equation}
\label{ineq:gamma}
0< c_0 \le \frac{1}{\Gamma(ap+b)} \le C_0,
\end{equation}
where for \(a>0\) and \(b>0\) the constants 
\(c_0\) and \(C_0\) depend only on \(a\), \(b\) 
and \(p_0\).
Thus, on the initial interval it suffices 
to estimate \(m_p\) instead of \(z_p\) in 
\eqref{geom:z_p}. 

To obtain the desired estimate, we first use 
Jensen's inequality to derive for every \(0<p'<p<p''\) 
the inequalities
\begin{equation}
\label{mom:interp}
\big(m^{1/p'}_{p'}\big)^p \le m_p 
\le \big(m_{p''}^{1/p''}\big)^{p}. 
\end{equation}
Taking \(p'=1\) and \(p''=p_0\) we obtain the 
bounds 
\begin{equation}
\label{geom:m_p}
c \,q^p \le m_p \le C\, Q^p
\end{equation}
for \(1\le p\le p_0\), with 
\(c=C=1\), \(q=m_1\) and \(Q=Q_0=\max\{1,m_{p_0}^{1/p_0}\}\).

{\em Step 1: Pure diffusion.} We take \(\eps=\frac{p_0-1}{2}\), 
use the bounds \eqref{K_eps1} and \eqref{K_eps2} in 
the moment inequalities \eqref{mom-ineq:Gauss}, 
\eqref{mom:Gauss} and estimate
\begin{equation}
\label{M_p}
S_p \le 2^{p+1} M_p\,, 
\quad\text{where}\quad
M_p = \max\limits_{1\le k \le k_p} 
\big\{m_k\, m_{p-k+\frac{1}{2}},m_{k+\frac{1}{2}}\, m_{p-k}\big\}. 
\end{equation}
We then obtain, for all \(p>1+\eps\), the inequalities
\begin{equation}
\label{mineq:m_p1}
2\,\mu\,(2p+1)\, m_{p-1} 
\le  m_{p+\frac{1}{2}} \le 
2\,K_\eps\,\mu\, p\,(2p+1)\, m_{p-1} + K_\eps 2^{p+1} M_p\,.
\end{equation}
Now we see that using \eqref{mineq:m_p1}
we can extend the bounds \eqref{geom:m_p}
(by augmenting the constants \(q\) and \(Q\) 
if necessary) to the interval 
\(\frac{3}{2}+\eps\le p\le p_0+\frac{1}{2}\). 
Using the interpolation inequality \eqref{mom:interp}
we can then extend the bounds \eqref{geom:m_p}
to all intermediate values \(p_0<p< p_0+\frac{1}{2}\). 

Further, by iterating inequalities 
\eqref{mineq:m_p1} we can cover the interval 
\(p_0\le p \le p_1\) by a fixed number of 
subintervals of length at most \(\frac{1}{2}\),
so that finally  inequalities 
\eqref{geom:m_p}, with the constants depending 
on \(m_0\), \(m_1\), \(p_0\) and \(m_{p_0}\) 
only, will be extended to the whole interval 
\(1\le p \le p_1\) . 

{\em Step 1: Diffusion with friction.} We argue as 
in the previous case and obtain using 
\eqref{mom-ineq:Gauss}, \eqref{mom:FP} the 
following  upper bounds for 
\(m_{p+\frac{1}{2}}\):
\begin{equation}\label{mineq:m_p2}
m_{p+\frac{1}{2}}
\le -2\,K_\eps\,\lambda\, p\, m_p 
+ 2\,K_\eps\,\mu\, p\,(2p+1)\, m_{p-1} 
+ K_\eps 2^{p+1} M_p\,,
\end{equation}
for all \(p>1+\eps\). Neglecting the non-positive friction term on 
the right-hand side yields the same upper 
bounds as in the pure diffusion case. On the 
other hand, the lower bound can be written in 
the form
\begin{equation}\label{mineq:m_p3}
2\,\lambda\, p\, m_p + m_{p+\frac{1}{2}} 
\ge 2\,\mu\, p\,(2p+1)\, m_{p-1} \,,
\end{equation}
which implies that for every \(p>1\) one of 
the following inequalities is true: 
\[
2\,\lambda\,p\, m_p
\ge \mu\, p\,(2p+1)\, m_{p-1}
\quad\text{or}\quad
m_{p+\frac{1}{2}} 
\ge \mu\, p\,(2p+1)\, m_{p-1}
\]
Combining the two inequalities and using the 
interpolation inequality \eqref{mom:interp}
in the second of the cases we obtain
\[\textstyle
m_p\ge\min\Big\{\,\frac{\mu}{\lambda}\,(p+\frac{1}{2})\,m_{p-1}\,,
\big(\,\mu\,p\,(2p+1)\,m_{p-1}\big)^{\frac{2p}{2p+1}}\Big\}. 
\]
This allows us to extend the lower bound \eqref{geom:m_p}
iteratively to the interval \(1\le p\le p_1\).
 
{\em Step 1: Self-similar solutions.} Using the moment
inequalities  \eqref{mom-ineq:Gauss}, \eqref{mom:selfsimi}
and arguing as above we obtain
\begin{equation}
\label{ineq:m_pS}
2 \kappa\, p\, m_p \le m_{p+\frac{1}{2}} 
\le 2 K_\eps \,\kappa\, p\, m_p
+ K_\eps\,2^{p+1}\,M_p\,,
\end{equation}
for all \(p\ge 1+\eps\). Using these bounds we extend 
\eqref{geom:m_p} to the interval \(1\le p\le p_1\) by the 
same iterative argument as in the pure diffusion case. 

We now pass to {\em Step 2} and use inequalities \eqref{mineq:z_p4},  
\eqref{mineq:z_p6} and \eqref{mineq:z_p7} to extend 
bounds \eqref{geom:z_p} to all \(p\ge1\) by an induction 
argument. The base of the induction is established by virtue 
of the bounds \eqref{geom:m_p} and \eqref{ineq:gamma} 
on the interval \(1\le p\le p_1\). We further verify 
the induction step separately in each of the three cases. 

{\em Step 2: Pure diffusion.} Our aim is to find the constants 
\(q\) and \(Q\) in such a way that for 
every \(n=1,2,3\dots\,\), the inequalities \eqref{geom:z_p} 
for \(1\le p \le p_1+\frac{n-1}{2}\) imply 
the same inequalities for \(p_1+\frac{n-1}{2} \le p 
\le p_1+\frac{n}{2}\). 
Thus, assuming \eqref{geom:z_p} for 
\(1 \le p \le p_1+\frac{n-1}{2}\) 
we use \eqref{mineq:z_p4} to find
\begin{equation*}
c_3\,\mu\,c\,q^{p-1} \le z_{p+\frac{1}{2}} 
\le C\Big( C_3\, K_{p_0}\, \mu\,Q^{-\frac{3}{2}} 
+ \frac{1}{2}\Big)\, \,Q^{p+\frac{1}{2}}.
\end{equation*}
Taking \(q\le (c_3\,\mu)^{\frac{2}{3}}\) and 
\(Q\ge(2C_3\,K_{p_0}\,\mu)^{\frac{2}{3}}\) 
we obtain the inequality
\[
c\,q^{p+\frac{1}{2}}\le z_{p+\frac{1}{2}} 
\le C\,Q^{p+\frac{1}{2}},
\] 
from which it follows that  \eqref{geom:z_p} is true 
for \(p_1+\frac{n-1}{2} \le p 
\le p_1+\frac{n}{2}\).

{\em Step 2: Diffusion with friction.} The upper bound 
case can be treated 
similarly to the previous one, with the difference that 
inequalities \eqref{mineq:z_p7} 
will allow us to increase \(p\) by {\em one} in each step, 
instead of {\em one half\(\,\)}, as in the pure diffusion 
case. Thus, for every \(n=1,2,3\dots\,\), we assume 
\eqref{geom:z_p} for \(1 \le p \le p_1+{n-1}\)
and obtain using  \eqref{mineq:z_p7},
\begin{equation*}
z_{p+1} \le  
\Big( \,\frac{C_3\,\mu}{C_5 \,\lambda} \,Q^{-1}
+ \frac{1}{2}\,\Big)\,C\,Q^{p+1}\,.
\end{equation*}
We now take  \(Q\ge\frac{2C_3\mu}{C_5\lambda}\) to obtain the 
inequalities
\[
z_{p+1} \le C\,Q^{p+1},
\]
which imply the upper bound \eqref{geom:z_p} for 
\(p_1+{n-1} \le p \le p_1+{n}\).

For the lower bound we see that assuming 
\eqref{geom:z_p} to be true for 
\(1 \le p \le p_1+\frac{n-1}{2}\), 
the inequalities \eqref{mineq:z_p6} imply 
that at least one of the inequalities 
\[
K_\eps\, C_5\, \lambda\, z_p \ge \frac{1}{2}\,\,c_3\,\mu \,c\,q^{p-1}
\quad
\text{or}
\quad
z_{p+\frac{1}{2}} \ge \frac{1}{2}\, c_3\,\mu \,c\,q^{p-1}
\]
is true. By choosing 
\(\displaystyle
q<\min\Big\{\big(\frac{1}{2}\, c_3\,\mu\big)^\frac{2}{3},
\frac{c_3\,\mu}{2K_\eps\,C_5\, \lambda}\Big\}\) we 
obtain \eqref{geom:z_p} for
\(p_1+\frac{n-1}{2} \le p \le p_1+\frac{n}{2}\).

{\em Step 2: Self-similar solutions.} We use inequalities 
\eqref{mineq:Shear4} and argue as in the pure diffusion 
case, assuming for every \(n=1,2,3\dots\,\) that  
\eqref{geom:z_p} holds for 
\(1\le p \le p_1+\frac{n-1}{2}\). 
We then find 
\[
c_5 \, \kappa\,c\, q^p \le z_{p+\frac{1}{2}} 
\le \Big( C_5 K_\eps \kappa\, Q^{-\frac{1}{2}} + \frac{1}{2} \,\Big) 
C\,Q^{p+\frac{1}{2}}.  
\]
Therefore, taking \(q<(c_5 \, \kappa)^2\) and 
 \(Q>(2C_5K_\eps\kappa)^2\) we obtain 
\eqref{geom:z_p} for 
\(p_1+\frac{n-1}{2}\le p \le p_1+\frac{n}{2}\). 

We can now complete the proof of Theorem~\ref{exp_bound:steady} 
by an induction argument.
\end{proof}

The above proof contains the proof of 
Theorem~\ref{exp_bound:steadyshear} as a special case: 
indeed the inequalities for the normalized moments 
in the shear flow case coincide with the upper 
inequalities for the case of self-similar solutions.
The result of 
Theorem~\ref{exp_bound:steadyshear} is weaker 
than in the latter case, since
we were not able to obtain suitable lower bounds 
for the moments in the shear flow problem.  

\section*{Concluding Remarks}

The estimates for the normalized moments that we 
established certainly deserve more attention that 
we were able to give them in the framework of the 
steady solutions to the kinetic equations. In 
fact, we hope to return to the problem of the 
time-evolution of the tails by the moment method
in a separate paper. Another promising direction
of study stems from the use of the integral 
bounds together with maximum principles for 
kinetic equations, in the form suggested by 
C. Villani \cite{V}. There are indications that 
such methods may yield more precise asymptotics
(in particular, pointwise upper bounds) for 
some cases of kinetic equations \cite{GaPaVi1}.

\section*{Acknowledgements}

The first author was supported by a grant from the 
Swedish Research Council (NFR). The second and third 
authors have been partially supported by NSF under grant 
DMS-0204568. Part of this research was supported
by the Pacific Institute for the Mathematical 
Sciences (PIMS) while the third author was a PIMS
post-doctoral fellow at the University of Victoria. 
The third author also acknowledges support by 
NSERC operating grant No. 7847. 
Support from the Institute for 
Computational Engineering and Sciences at the 
University of Texas at Austin is also gratefully 
acknowledged.

\bibliographystyle{hsiam}
\bibliography{gran_m5}

\end{document}